\documentclass[twocolumn, twocolappendix]{aastex631}
\received{\today}
\shorttitle{NEO Follow-up in Era of LSST}
\graphicspath{{figures/}}

\usepackage{lipsum}
\usepackage{physics}
\usepackage{multirow}
\usepackage{xspace}
\usepackage{natbib}
\usepackage{fontawesome5}
\usepackage{xcolor}
\usepackage{wrapfig}
\usepackage[figuresright]{rotating}

\usepackage[hang,flushmargin]{footmisc}

\newcommand{\dig}{\texttt{digest2}\xspace}
\newcommand{\sss}{S3M\xspace}
\newcommand{\mpco}{MPCORB\xspace}

\newcommand{\nightlyTrafficBase}{100\xspace}
\newcommand{\nightlyTrafficAllMags}{129\xspace}
\newcommand{\nightlyPurityBase}{8.3\%\xspace}
\newcommand{\digestThresholdPurity}{5.4\%\xspace}
\newcommand{\digestMaxPurity}{64.1\%\xspace}
\newcommand{\npernightAlg}{64\xspace}
\newcommand{\purityAlg}{8.4\xspace}
\newcommand{\purityAlgRaw}{5\xspace}
\newcommand{\neoLostAlg}{189\xspace}
\newcommand{\efficiencyAlg}{68\xspace}
\newcommand{\thresholdAlg}{0.8\xspace}

\newcommand{\edtom}[1]{#1}

\makeatletter
\newcommand{\unit}[1]{%
    \,\mathrm{#1}\checknextarg}
\newcommand{\checknextarg}{\@ifnextchar\bgroup{\gobblenextarg}{}}
\newcommand{\gobblenextarg}[1]{\,\mathrm{#1}\@ifnextchar\bgroup{\gobblenextarg}{}}
\makeatother

\begin{document}

\title{Expected Impact of Rubin Observatory LSST on NEO Follow-up}

\newcommand{\UW}{DiRAC Institute and the Department of Astronomy, University of Washington, Seattle, WA, 98195}

\newcommand{\UIUC}{Department of Aerospace Engineering, University of Illinois Urbana-Champaign, Urbana, IL, 61801}

\author[0000-0001-6147-5761]{Tom Wagg}
\affiliation{\UW}

\author[0000-0003-1996-9252]{Mario Juric}
\affiliation{\UW}

\author[0000-0003-2874-6464]{Peter Yoachim}
\affiliation{\UW}

\author[0009-0005-5452-0671]{Jake Kurlander}
\affiliation{\UW}

\author[0000-0002-0672-5104]{Sam Cornwall}
\affiliation{\UIUC}

\author[0000-0001-5820-3925]{Joachim Moeyens}
\affiliation{\UW}

\author[0000-0002-1398-6302]{Siegfried Eggl}
\affiliation{\UIUC}

\author[0000-0001-5916-0031]{R. Lynne Jones}
\affiliation{\UW}

\author[0009-0007-9898-7358]{Peter Birtwhistle}
\affiliation{Great Shefford Observatory, Hungerford, England}

\correspondingauthor{Tom Wagg}
\email{tomjwagg@gmail.com}

\begin{abstract}
    We simulate and analyse the contribution of the Rubin Observatory Legacy Survey of Space and Time (LSST) to the rate of discovery of Near Earth Object (NEO) candidates, their submission rates to the NEO Confirmation page (NEOCP), and the resulting demands on the worldwide NEO follow-up observation system. We find that, when using current NEOCP listing criteria, Rubin will typically contribute ${\sim}$\nightlyTrafficAllMags new objects to the NEOCP each night in the first year, an increase of ${\sim}8$x relative to present day. Only \nightlyPurityBase{} of the objects listed for follow-up will be NEOs, with the primary contaminant being a background of yet undiscovered, faint, main belt asteroids (MBAs). We consider follow-up prioritisation strategies to lessen the impact on the NEO follow-up system. We develop an algorithm that predicts (with $\efficiencyAlg{}\%$ accuracy) whether Rubin itself will self recover any given tracklet; external follow-up of such candidates can be de-prioritised. With this algorithm enabled, the follow-up list would be reduced to $\npernightAlg{}$ NEO candidates per night (with ${\sim}\purityAlg{}\%$ purity). We propose additional criteria based on trailing, apparent magnitude, and ecliptic latitude to further prioritise follow-up. We hope observation planners and brokers will adopt some of these open-source algorithms, enabling the follow-up community to effectively keep up with the NEOCP in the early years of LSST.
    
\end{abstract}


\section{Introduction} \label{sec:intro}
Near-Earth Objects (NEOs) are asteroids and comets that have a perihelion distance less than $1.3 \unit{au}$. It is estimated that approximately one fifth of this population passes close enough to Earth that small perturbations in their orbit may lead to intersections with the Earth's orbit and potential collisions \citep[e.g.][]{Jones+2018}. A subset of these objects are known as Potentially Hazardous Asteroids (PHAs), these objects are defined as being at least 140m in diameter that pass within 0.05au of the Earth\footnote{\url{https://cneos.jpl.nasa.gov/about/neo_groups.html}}. PHAs are large enough to make it through the Earth's atmosphere and still cause continent scale damage through impact. Given the threat posed by these objects, a world-wide effort\footnote{E.g.\,\url{https://www.unoosa.org/oosa/en/ourwork/topics/neos/index.html}} has been ongoing to catalogue and determine the orbits and sizes of NEOs including identifying any posing a hazard to the Earth.

The Minor Planet Center maintains a catalogue of known NEOs and their orbits\footnote{\url{https://www.minorplanetcenter.net/iau/MPCORB/NEA.txt}}, as well as the NEO confirmation page (NEOCP\footnote{\url{https://www.minorplanetcenter.net/iau/NEO/toconfirm_tabular.html}}). The NEOCP is a continously updated web page listing newly discovered NEO candidates that should be prioritised for additional observations by the NEO follow-up community. These follow-up observations contribute additional astrometric observations necessary to more accurately determine the orbit of the candidate, as well as photometry to constrain its size. An object is only listed on the NEOCP when it has a high probability of being an NEO. This probability is quantified using the \dig{} code \citep{Keys+2019}. \dig{} assigns a score between 0 and 100 based on potential orbits that fit the observations and only objects with a score of 65 or more are listed on the page. \edtom{Currently, on average between ${\sim}5$-$27$ objects are added to the NEOCP on each night, varying as a function of lunation and season (see Appendix~\ref{app:neocp} for more details).}

The Rubin Observatory Legacy Survey of Space and Time \citep[LSST,][]{Ivezic+2019} will rapidly increase the rate at which NEO candidates are identified and reported to the NEOCP. \citet{Jones+2018} showed that at the end of the 10-year LSST baseline survey the completeness of NEOs with an absolute magnitude of $H \le 22$ would be 73\%. Most of these objects will be discovered using ``tracklet linking'': a computational technique where at least three pairs of observations (``tracklets'') observed over a 15-night period are identified as belonging to the same object (\citealp{Juric+2017}; Heinze et al., in prep). The orbits of objects discovered with this technique will typically be reasonably well known, and in need of no immediate follow-up. However, this tracklet linking comes at a cost: the object is not identified as interesting until the third tracklet is imaged -- at best, two nights after the first observation or, at worst, nearly two weeks later. This means that potentially interesting (or hazardous) objects may be missed until it is too late to observe (or react to) them.

A more traditional discovery technique would be to take enough back-to-back images so high-confidence tracklets can be built with three or more observations and immediately reported for follow-up. The LSST cannot do that, as it would reduce the efficiency of other science areas the data are to support\footnote{For example, the Dark Energy science case prioritises obtaining a uniform scan of the entire visible sky over frequently revisiting any given area.}. However, in a smaller area of the sky (e.g., where there are adjacent field overlaps), the LSST {\em will} serendipitously produce 3+ observation tracklets. Such tracklets could be immediately identified and, assuming they meet the \dig{} score criteria, submitted to the MPC and included on the NEOCP. \edtom{Given the scale, this process will be automated and typically involve no human vetting.}

The aim of this paper is to quantify the impact of Rubin on the NEO follow-up community and consider possible strategies to mitigate this impact. We performed mock LSST observations and used \dig{} to assess the number of objects observed by LSST that could be submitted to the NEOCP. We present an algorithm for predicting whether LSST will later re-detect an object given a single night of observations (therefore making community follow-up unnecessary). We apply this algorithm to the mock observations and quantify by how much we could reduce the number of objects requiring community follow-up.

The paper is structured as follows. In Section~\ref{sec:current_impact}, we estimate the impact that LSST will have on the NEOCP when applying current submission criteria. We describe the methods for simulating LSST observations, calculating \dig{} scores and present our findings for the total number and type of objects submitted to the NEOCP by LSST. In Section~\ref{sec:mitigation}, we present an algorithm for prediction of LSST self-followup probabilities and assess how well its application reduces the impact on the NEOCP. Based on these results, in Section~\ref{sec:further_strats} we discuss further mitigation strategies that may be needed and make recommendations for how the follow-up of Rubin-discovered objects could be prioritised. We summarise our conclusions in Section~\ref{sec:conclusion}. All code needed to reproduce results and figures, as well as the implementation of our algorithms, is available in a GitHub repository\footnote{\url{https://github.com/TomWagg/lsst-neocp-predictions/}}.

\section{The impact of Rubin observations on the NEOCP}\label{sec:current_impact}
To make predictions for the NEOCP in the era of LSST, we perform simulated observations of a `hybrid' catalogue of solar system objects that self-consistently combines known and simulated populations. We then use the \dig{} code to calculate NEO scores for each object and use these values to make predictions for the NEOCP.

\subsection{Simulated observations}\label{sec:sim_obs}
We start by developing a `hybrid' solar system object catalogue to investigate the effect of LSST sources on the NEOCP. This catalogue contains real objects (from MPCORB, \citealt{mpcorb}) and synthetic objects (from the Pan-STARRS Synthetic Solar System Model, \sss, \citealt{Grav+2011}). These catalogues are combined such that they retain the same overall distributions in position, velocity and absolute magnitude as found in the purely synthetic catalogue (see Appendix~\ref{app:hybrid}). This makes it possible to exclude already discovered objects from the simulated lists of objects the LSST will find.

\subsubsection{Rescaling the \sss model}\label{sec:rescaling-s3m}

\begin{figure*}
    \centering
    \includegraphics[width=\textwidth]{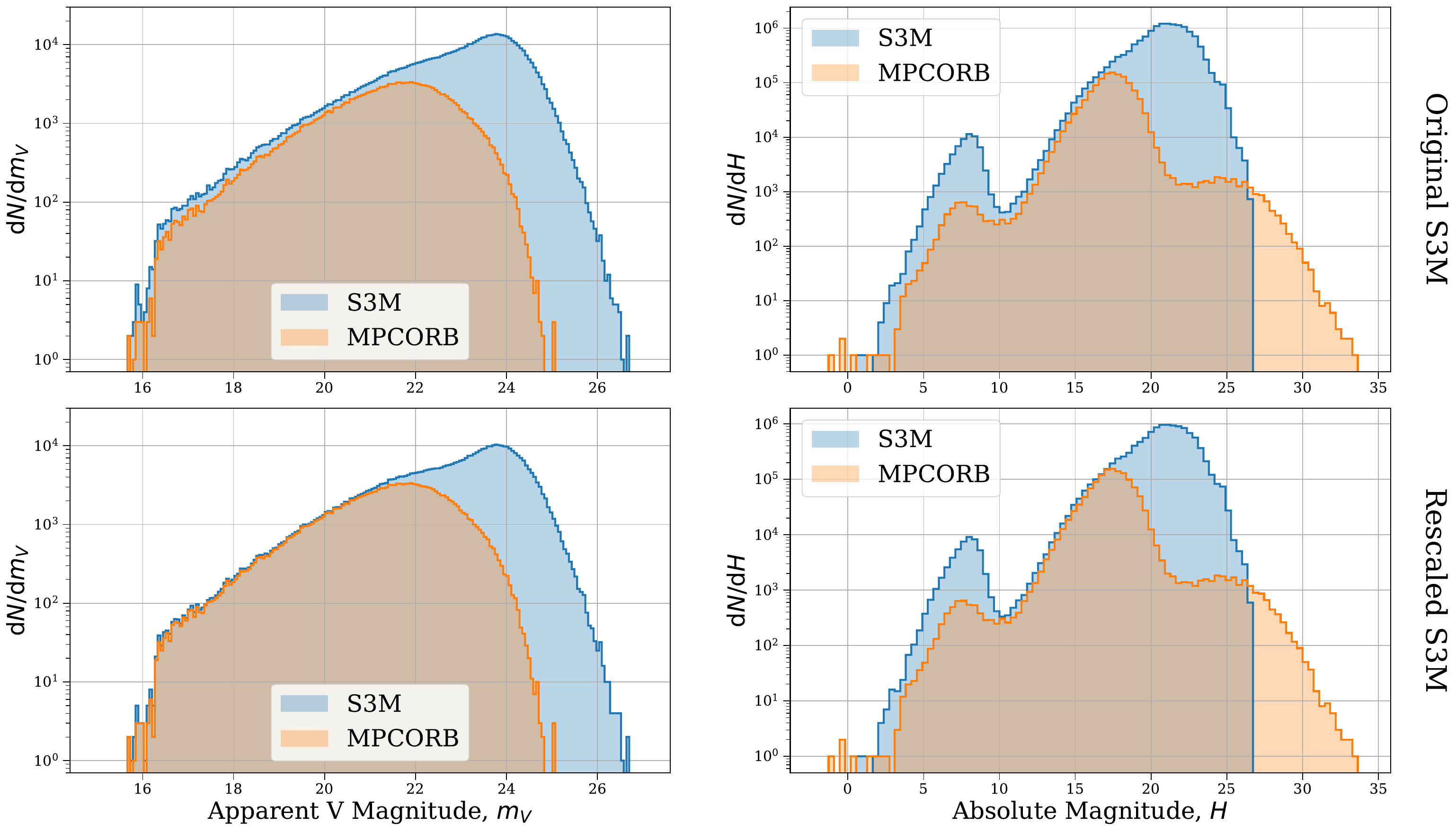}
    \caption{A demonstration of our rescaling of the synthetic solar system catalogue S3M to match MPCORB observations at bright magnitudes. All panels compare known objects (from MPCORB) with synthetic objects (from S3M). The left column shows the distribution of apparent magnitude of observations of asteroids for a single representative night of LSST. The top row shows the original S3M distributions and the bottom row shows the distribution after applying our rescaling.}
    \label{fig:rescaling}
\end{figure*}

\edtom{The \sss model, which was calibrated to observations available in early 2011, overestimates the number of solar system objects seen at present. Compared to MPCORB, we find the \sss predicts around 25\% more objects from $m_V = 17$ to $m_V = 20.5$, a range where the sample is expected to be reasonably complete. For example, the limiting magnitudes of the Catalina Sky Survey\footnote{\url{https://catalina.lpl.arizona.edu/about/facilities}} and PanSTARRS \citep{Chambers+2016:2016arXiv161205560C}, the primary discoverers of NEOs, are both fainter than $m_V = 20.5$.}

\edtom{The excess of objects in S3M can be seen in the upper left panel of Figure~\ref{fig:rescaling}, in which we plot the distribution of apparent magnitudes for objects in S3M and MPCORB for a representative night of observations. Similarly, in the upper right panel we plot the distributions of absolute magnitude. The excess is again visible between $H = 10$ and $H = 17$ (the excess for $H < 10$ is from trans-Neptunian objects).}

\edtom{Given the excess of objects in S3M, we rescaled the simulation to match with MPCORB. We rescale by uniformly randomly retaining a fraction $f$ of objects in the catalogue. We searched over a range of choices of $f \in [0.75, 0.85]$ in increments of $0.01$ and found that $f = 0.80$ resulted in the closest match between S3M and MPCORB at $m_V < 20.5$. The lower panels of Figure~\ref{fig:rescaling} show the distributions after rescaling S3M and demonstrate that rescaling results in a strong agreement between S3M and MPCORB below $m_V = 20.5$.}

\edtom{The rescaling of S3M is a useful solution to enable this work, but also points to a need for an updated synthetic solar system catalogue. This is something that the LSST itself will enable in the next few years, and as such is beyond the scope of this paper. We additionally note that most predictions for LSST solar system detections to date have relied upon S3M; it's therefore possible the rates have been systematically overestimated.}

\subsubsection{Observing strategy}

\begin{figure}[tb]
    \centering
    \includegraphics[width=\columnwidth]{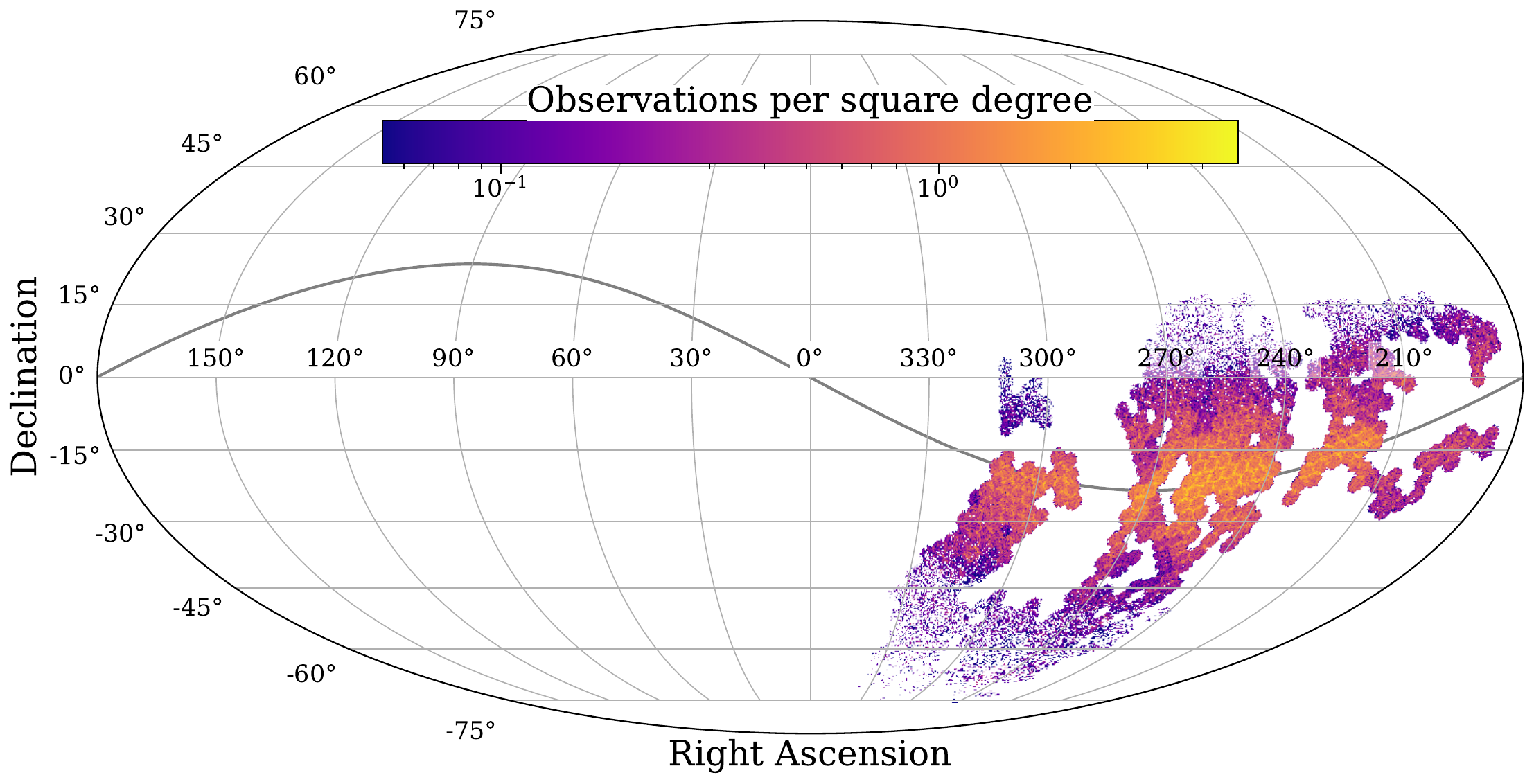}
    \caption{An example of asteroids detected in a single night of LSST observations. This example has ${\sim}350,000$ observed asteroids, of which ${\sim}1000$ are NEOs. Grey curve indicates the ecliptic plane.}
    \label{fig:observations_per_night}
\end{figure}

We perform mock LSST observations on the hybrid catalogue that we constructed. We use the ``Baseline v3.3'' 10 year scheduler simulation strategy \citep{Naghib+2019, Cornwall+2020}. These observations account for both scheduled and unscheduled downtime and simulate the current baseline observing strategy that will be followed by LSST. The resulting simulations span nearly 3600 days, and consist of near a billion observations. An example of asteroids detected in a single night of observations is shown in Figure~\ref{fig:observations_per_night}.

\subsection{Building and Scoring Tracklets}\label{sec:digest2_score}
We begin our analysis by selecting observations from the simulated LSST dataset that correspond to tracklets. For a tracklet to be built, we require it to satisfy the following criteria:
\begin{enumerate}
    \item \textbf{Number of observations:} We consider only objects which have at least 3 observations on a given night (we also consider more stringent criteria of 4+ observations - see Section~\ref{sec:traffic_basic}).
    \item \textbf{Maximum time separation:} We set the maximum time between observations to 90 minutes. Thus we only allow tracklets that have at least one pair of observations that occur within 90 minutes.
    \item \textbf{Minimum arc length:}\footnote{\edtom{Note that this refers to 1 arcsecond across multiple observations, not in a single exposure.}} We ensure that each tracklet is at least 1 arcsecond in length (corresponding to ${\sim}5$ pixels on Rubin's LSSTCam camera). This ensures that the motion vector of the tracklet can be determined.
\end{enumerate}
The motivation behind these cuts is to ensure that the tracklet constrains the on-sky motion of the object sufficiently well, so that its position at a later time can be easily extrapolated. With fewer observations or shorter tracklets, many different orbits could reproduce the same motion on the sky. Moreover, observations that are separated too significantly in time may be spurious linkages, where observations of multiple objects are incorrectly assumed to be of the same source.

\begin{figure}[tb]
    \centering
    \includegraphics[width=\columnwidth]{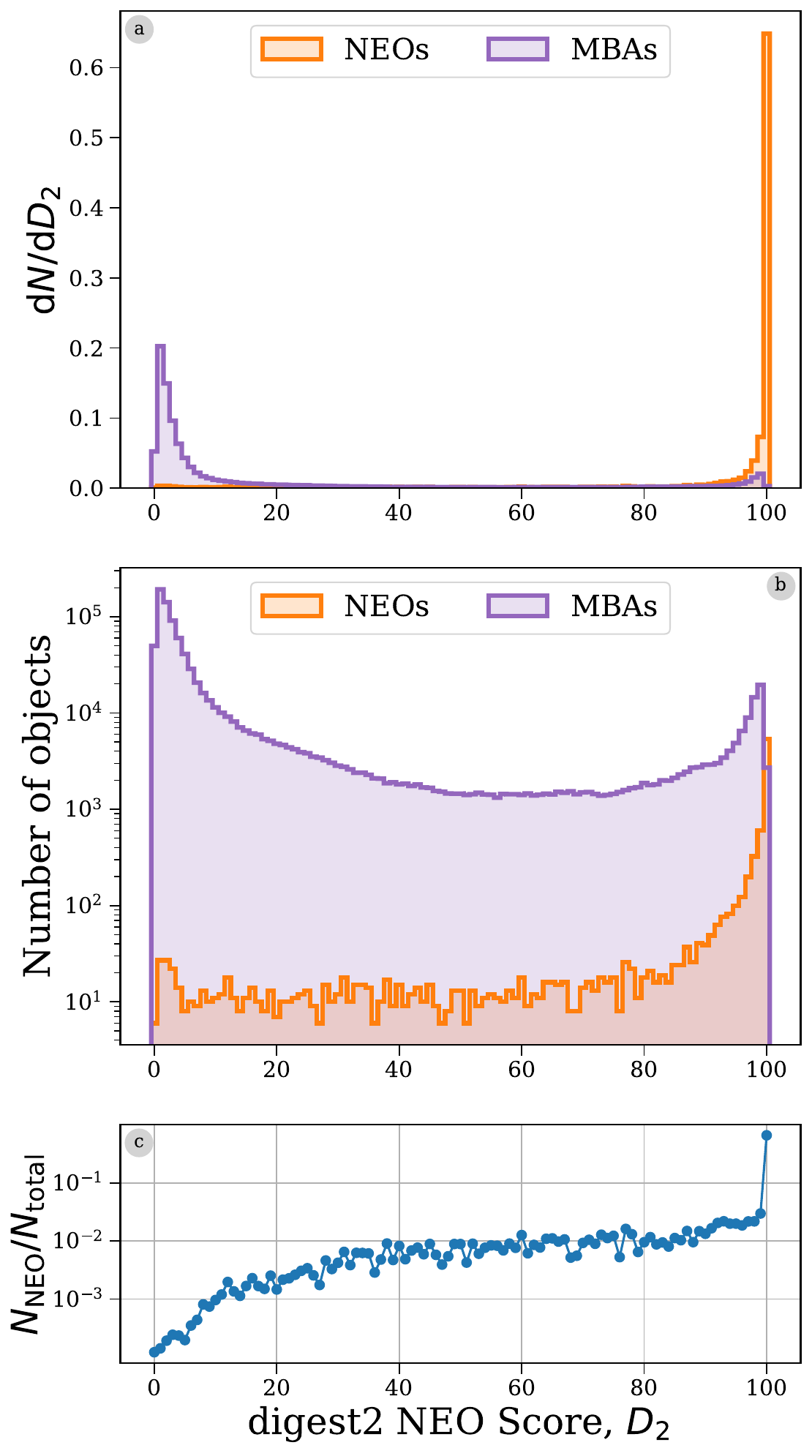}
    \caption{The \dig{} score cannot adequately distinguish NEOs in the presence of a large MBA background. \dig{} scores for all NEOs and MBAs observed in the first year of our simulated LSST observations. \textbf{(a)} normalised histograms of \dig{} scores, \textbf{(b)} the same histograms un-normalised \textbf{(c)} NEO histogram divided by the sum of the histograms in (b). Note that the latter two panels are on a logarithmic scale.}
    \label{fig:digest2_example}
\end{figure}

A single tracklet in itself does not determine the orbit. Tracklets do place constraints on the direction and rate of motion which can be used -- when compared to the known populations of objects -- to determine the probability of the object having a particular orbit. These can then be marginalized over classes of interest to score the chance of an object belonging to any given class of objects. The main criterion the Minor Planet Center uses to place an object on the NEOCP is an NEO \dig{} score of at least 65. This score, ranging from 0 to 100, quantifies the probability that the object is an NEO and is calculated using the \dig{} code \citep{Keys+2019}.

At its core, \dig{} compares a simulated catalogue of solar system objects to observed tracklets to estimate the probability that an object is an NEO. \dig{} bins simulated objects into 15 different orbit classes, using bins of perihelion, eccentricity, inclination and absolute magnitude. Then, for each observed tracklet, \dig{} samples a series of possible distances and radial velocities and uses those values to estimate possible orbits of the object. These orbits are binned and assigned a class based on the bin they are assigned. The NEO score is then estimated as a fraction of the orbits that are classified as NEOs. For a more exhaustive description of \dig{}, see \citet{Keys+2019}. We use \dig{} to calculate the NEO score of each tracklet in our sample. 

The distribution of \dig{} scores for the NEOs and MBAs in the first year of observations is shown in Figure~\ref{fig:digest2_example}. As one would expect most NEOs have scores around 100, whilst most MBAs have scores around 0. However, we can already see that due to the volume of MBA observations, the \dig{} score alone will not result in a high-purity sample of NEOs candidates. \edtom{In particular, only \digestThresholdPurity of objects assigned a score of 65 or more are actually NEOs (Figure~\ref{fig:digest2_example}c). Even requiring a score of at least 90 only increases this to $8.1\%$, only when requiring a score of exactly 100 does the probability significantly increase to \digestMaxPurity (Figure~\ref{fig:digest2_example}c)}. We discuss this further in Section~\ref{sec:digest2_improve}.

\subsection{Rubin NEOCP Submissions: Rates and Purity}\label{sec:traffic_basic}

\begin{figure*}[htb]
    \centering
    \includegraphics[width=\textwidth]{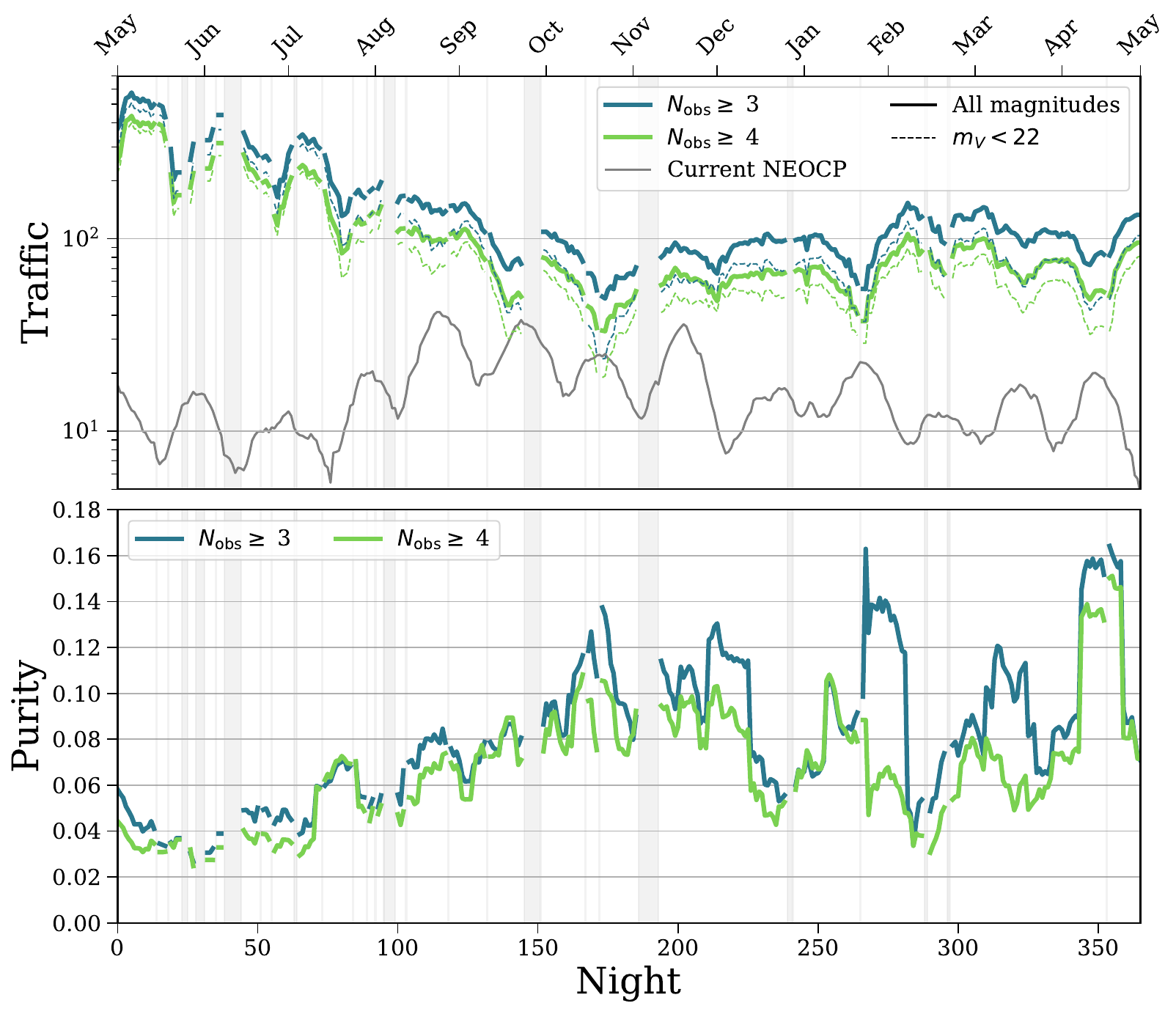}
    \caption{Traffic (number of objects sent) and purity (fraction of objects sent that are NEOs) of the NEOCP during the first year of LSST if every observation that qualifies for submission and with an apparent magnitude of $m < 22$ is submitted. Each line is plotted using a rolling window of 10 days to smooth stochastic effects. Different colours correspond to different constraints on the number of observations for a tracklet to be submitted. In the top panel, solid lines indicate traffic for any magnitude, whilst dashed lines are limited to $m_V < 22$). Nights on which no observations were taken are highlighted with grey areas.}
    \label{fig:neocp_traffic}
\end{figure*}


Figure~\ref{fig:neocp_traffic} shows the impact that LSST submissions would have on the NEOCP if \textit{every} new tracklet that met the \dig{}~$\geq65$ 
were submitted. We show the rates both irrespective of magnitude, and with a $m<22$ ``bright'' object cut (discussed further in Section~\ref{sec:priority_mag}). The top panel shows the number of new NEO candidates added to the NEOCP each night\footnote{Crucially, we assume that if an object is re-observed again in the following days, it will be successfully linked to the initially submitted tracklet and not counted as an unknown candidate again.}, whilst the bottom panel shows the purity of the submissions: the fraction of candidates that are actually NEOs. We additionally show the difference of requiring an additional observation beyond the original choice of 3 as discussed in Section~\ref{sec:digest2_score}).

\edtom{Initially, LSST rates exceed several hundred submissions per night. After around 150 nights the rate of submissions reaches a steady state, at around 95 submissions per night. On average we find that ${\sim}$\nightlyTrafficAllMags objects would be submitted per night over the first year. One can also note short term variations as a result of the lunar cycle. For comparison, we show the archival traffic on the NEOCP for May 2023--2024\footnote{Note that the difference in year means the lunar phases are offset.} with a grey line \citep[see Appendix~\ref{app:neocp}]{neocp-archives}. As suspected, we predict LSST contributions will significantly exceed the current rates. Even when applying a magnitude cut of $m_V < 22$ the rate remains high at ${\sim}\nightlyTrafficBase$ submissions per night. The relatively small decrease is due to the fact that a \dig{} threshold of 65 is already effective at limiting contamination from faint MBAs (see Section~\ref{sec:priority_mag})}.

\edtom{The purity of submitted candidates is impacted by abundant, hitherto unidentified, MBA observations contaminating the sample. As these are discovered and can be removed from the sample, the purity starts increasing until it levels off some $\sim$5 months into the survey\footnote{At that point, the unknown MBAs become concentrated in the new sky rising in the East, thus reducing the overall contamination rate.}. Beyond short term lunar variations, the purity otherwise remains fairly consistent throughout the rest of the year, with an average of around \nightlyPurityBase. Should this sample be followed up without further filtering, a significant amount of follow-up time would be spent re-observing MBAs masquerading as NEOs.}

\edtom{Finally, we note that these predictions are model dependent. For example, \citet{Granvik+2018:2018Icar..312..181G} presents a different model for NEOs. This model predicts almost a factor of 2 more bright NEOs than S3M \citep{Granvik+2018:2018Icar..312..181G, Grav+2011}. Therefore, if one adopts this model, our predictions would change proportionally: in this case increasing the purity but leaving the traffic (which is dominated by MBAs) relatively unchanged.}



\section{Mitigating the impact of LSST}\label{sec:mitigation}

Given that Rubin repeatedly re-observes large swaths of the sky, it will recover many potential NEOs on its own, providing sufficient astrometry for accurate orbit determination without external follow-up. If we could predict which objects will be followed up by LSST itself, this would reduce the load on the follow-up system and allow the community to focus on ones that truly \emph{require} external follow-up to be designated.

\subsection{The best-case scenario: perfect prediction of self-follow-up}\label{sec:no_LSST_detections}

Rubin requires the pipelines link 95\% of objects observed on at least 3 separate nights within a 15 day window, each with at least 2 observations separated by at most 90 minutes \citep{oss}. As these pipelines are still under development, we compute how many objects Rubin can link on its own by using a python package \texttt{difi} \citep{difi}. This package emulates a ``perfect linker'', identifying tracklets that Rubin pipelines will be able to link and compute orbits for. We've confirmed that the development version of Rubin linking pipelines are very close to the performance emulated by \texttt{difi} (Ari Heinze, priv. comm.).

\edtom{Using \texttt{difi}, we find that removing objects that would not require follow-up decreases the average NEO candidate rates for the first year by around a factor of 2, to 55 follow-up candidates per night. Additionally the average purity slightly increases to 10.3\% over the first year.}


\subsection{Estimating the probability of self-follow-up}\label{sec:pred_alg}
Motivated by these potential improvements, we look at developing an algorithm that assesses whether an object can be detected without external follow-up. Note that this will be an inherently imprecise, probabilistic, estimate: the LSST schedule is dynamic, the weather cannot be predicted exactly, and the NEO candidate itself could take many different paths on the sky within the allowable observational errors.

The technique we adopt is conceptually simple. Given a single tracklet, one can compute the ensemble of ranges and radial velocities $(D, \dot{D})$ of NEOs consistent with that tracklet. We can compute the {\em expected} sequence of Rubin observations for the next month, assuming clear weather\footnote{More sophisticated weather forecasting could be deployed once the survey is ongoing.}. Finally, we cross-correlate the two, and compute the fraction of the NEO ensemble that would be recovered.

The details and the implementation are as follows. In order to determine the location of an object on subsequent nights one needs to estimate its orbit. Each object on a given night consists of at least two observations, $O$, that have the form
\begin{equation}
    O = \{ \alpha, \delta, t \}
\end{equation}
where $\alpha$ is the right ascension, $\delta$ is the declination and $t$ is the time of observation. We determine the proper motion of the object on the sky, ($\dot{\alpha}$, $\dot{\delta}$), by calculating the change in position between the two observations using \texttt{Astropy SkyCoord} and dividing by the time between observations. This determines 4 of the 6 orbital elements, but the topocentric distance, $D$, and radial velocity, $\dot{D}$, of the object are unconstrained.

\begin{figure}[tb]
    \centering
    \includegraphics[width=\columnwidth]{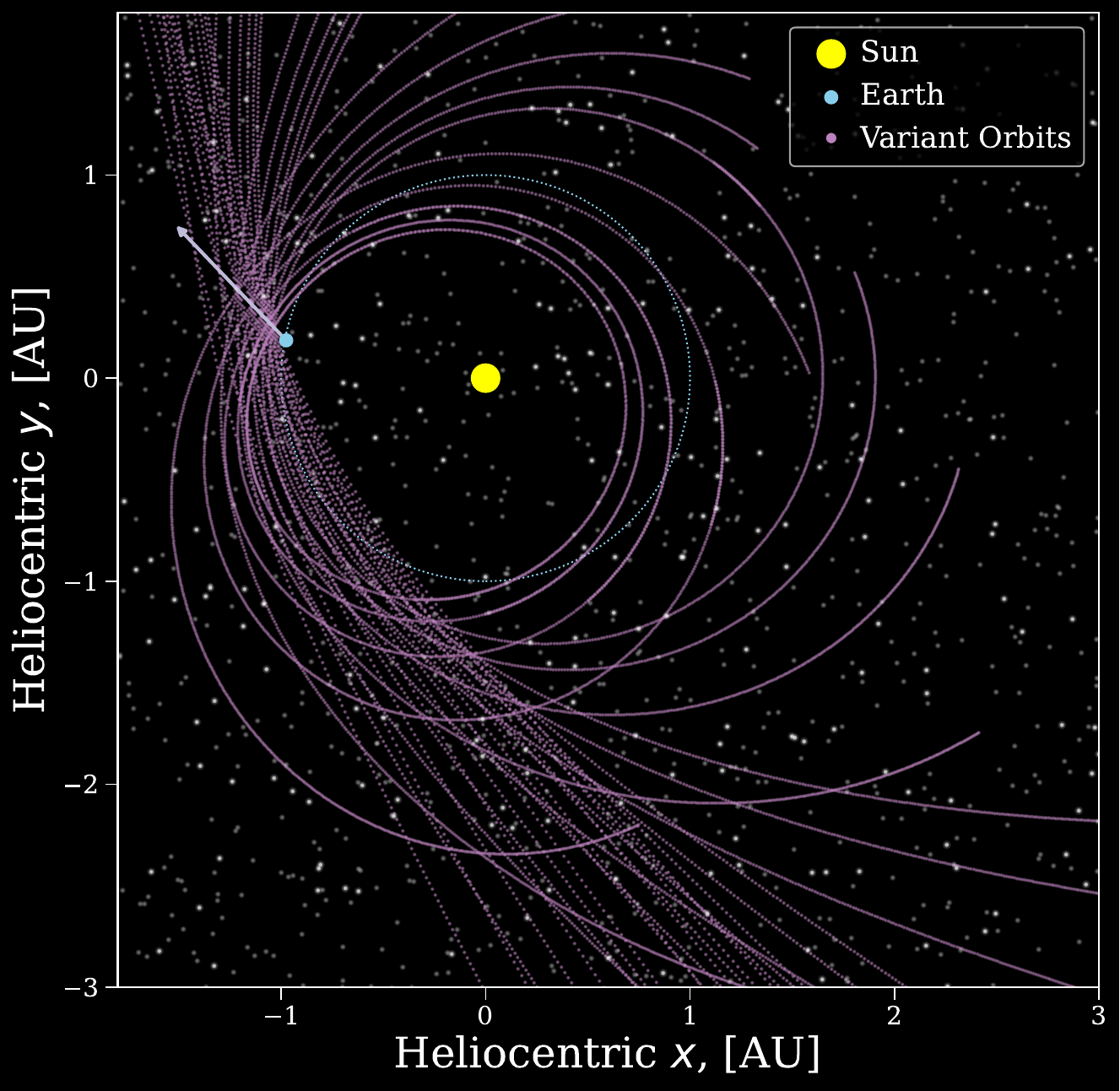}
    \caption{Variant orbits computed for an example NEO in our sample based on a single night of mock LSST observations. The white arrow indicates the initial sight-line for the observation. The blue dotted line indicates the orbit of the Earth. Background stars are for illustrative purposes only.}
    \label{fig:orbits}
\end{figure}

We draw a sample of $D$ uniformly in log-space between $[0.1, 10] \unit{au}$ and a sample of $\dot{D}$ uniformly between $[-50, 10] \unit{km}{s^{-1}}$ to create a grid of $(D, \dot{D})$ \citep[e.g.][]{Virtanen+2001:2001Icar..154..412V}. Combining these with the measured $(\alpha, \delta, \dot{\alpha}, \dot{\delta})$ values (and adding $0.1^{\prime\prime}$ of scatter to the measured on-sky positions to account for the detector uncertainty), we create a series of possible variant orbits for the object. We use the functionality in the \texttt{THOR} package \citep{Moeyens+2021} to handle the orbital dynamics and celestial mechanics (including light-time travel corrections). Since we are only interested in NEOs, we mask out any orbits that have a perihelion distance of greater than $1.3 \unit{au}$. To demonstrate the different possible orbits that we could infer from a single observation, we plot a subset of the variant orbits obtained for an NEO observed in our mock observations in Figure~\ref{fig:orbits}.

Additionally, we estimate the absolute magnitude of the object for each orbit using the $HG$-system \citep{mpc_h_g}. This is needed to assess whether the object will be bright enough to detect in subsequent observations. To convert to LSST band magnitudes, we assume a C-type asteroid and mean colours adopted from \citep{Jones+2018}. For the slope parameter, we assume the customary $G=0.15$ \citep{mpc_h_g}.

For each night in the first year, we use \texttt{rubin\_sim} to create a predicted schedule for the following 14 nights, our assumed detection window \citep{rubin_sim}. These predicted schedules represent an estimate of where LSST will look next and as such account for scheduled downtime, but do not include unscheduled downtime or poor weather conditions. This means our predicted schedule represents the best-case scenario; once Rubin is in commissioning and additional information about the weather becomes available it will be straightforward to incorporate it into the analysis.

\begin{figure*}[htb]
    \centering
    \includegraphics[width=\textwidth]{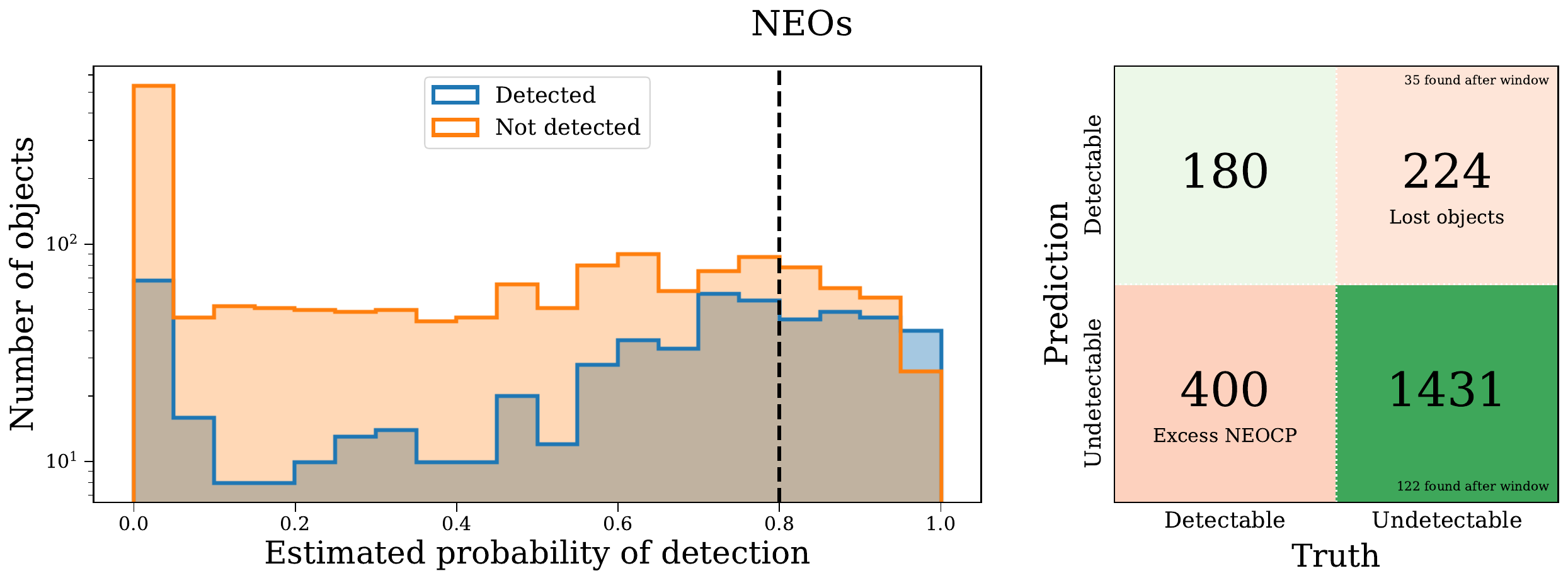}
    \includegraphics[width=\textwidth]{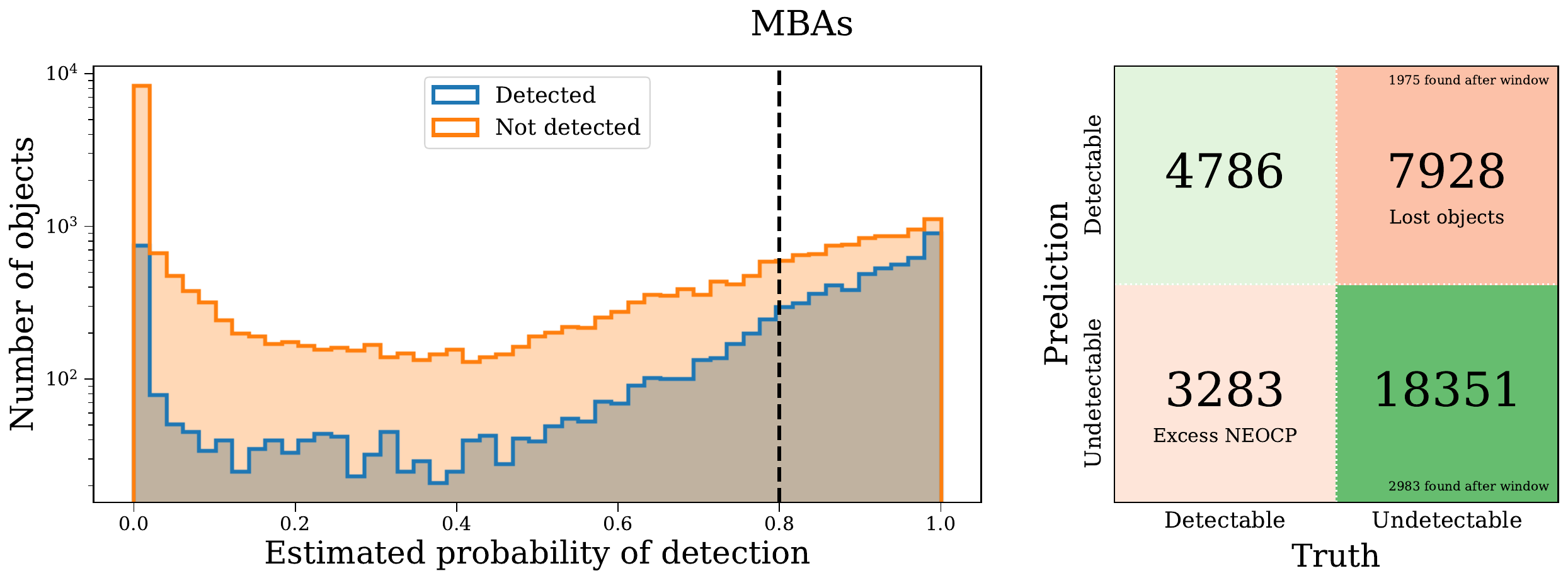}
    \caption{A demonstration of the prediction algorithm described in Section~\ref{sec:pred_alg} using 1 year of simulated observations for both NEOs (top) and MBAs (bottom). \textbf{Left:} Probability that an object will be detected within 15 nights, split into a population of objects that are \textit{actually} detected within 15 nights in the simulated observations and a population that are not. The black dashed line indicates our threshold of $\thresholdAlg{}$ for submission to the NEOCP. \textbf{Right:} A contingency matrix representing the algorithm's ability to predict the detectability of an object. The truth columns correspond to whether an object would be successfully followed-up by Rubin and the prediction columns correspond to whether the probability predicted by our algorithm surpassed a threshold of 0.8. Lower right quadrants are labelled as ``excess NEOCP'' given that the objects would be unnecessarily be prioritised for external follow-up. Upper right quadrants are labelled ``lost objects'' since they would be de-prioritised for external follow-up, but not receive adequate follow-up from Rubin. Small annotated values indicate the number of objects that would be followed-up by Rubin by the end of the first year of LSST, but \textit{after} the initial 15 night window.}
    \label{fig:contingency}
\end{figure*}

Next, using \texttt{OpenOrb}, we produce ephemerides for all variant orbits at times of each visit in the following 14 nights of the predicted schedule \citep{Granvik+2009}. For each orbit and each visit, we check whether the object is within the LSST camera footprint, and whether it is bright enough to be detected using Rubin's canonical $5\sigma$ cutoff \edtom{\citep[][Table 1]{Ivezic+2019}}. If both of these criteria are met then we assume an observation has been made.

\edtom{Finally, we determine the fraction of variant orbits whose tracklets satisfy LSST linking requirements, such that they have} at least 2 observations on at least 3 nights in a 15 day window, with each having a minimum arc length of 1 arcsecond and maximum time separation of not more than 90 minutes \citep{oss}. We also account for prior observations in this calculation. \edtom{As an example, consider an object that was observed on night 10 and had previously been observed on night 8. In this case, we would account for the prior observation, and so even a single additional future tracklet in the linking window (on, for example, night 14) is sufficient to complete the criterion of 3 nights for linkage. Following this procedure, the overall probability of the object being detected by LSST is then a simple fraction of variant orbits that were successfully linked.}

\subsection{Avoiding the follow-up of likely-to-be-linked objects}\label{sec:using_alg}
We now examine the effect of applying the LSST detection probability algorithm to reduce the load on the NEOCP. Figure~\ref{fig:contingency} shows the results for a year of simulated observations for both the NEOs and MBAs, including only objects that we would potentially submit to the MPC (at least 3 observations in the initial tracklet and with an apparent magnitude of at most $22$). The histograms show the estimated probability of detection by LSST for each object, split into a population of objects that are \textit{actually} detected within 15 nights in our simulated observations and a population that are not. On average\edtom{, when compared to the true outcome calculated by \texttt{difi},} the algorithm predicted the correct outcome approximately $\efficiencyAlg{}\%$ of the time.

We use a threshold of $\thresholdAlg{}$ for deciding whether an object will be detected by LSST. We found that this value is a good balance between the number of NEOs that are lost and the number of candidates that need to be followed up, whilst at the same time maximising the fraction of objects that are truly NEOs.

In the right side of Figure~\ref{fig:contingency}, we show contingency matrices for both NEOs and MBAs when applying this threshold. The lower left quadrants give a count of the number of objects that will be sent to the NEOCP needlessly as LSST will detect them without follow-up. The upper right quadrants total the number of objects that will be lost, since LSST will not detect them within the given detection window but they would also not be sent to the NEOCP. We additionally annotate the number of objects that would be found after the detection window and note this reduces the overall number of lost objects.

Overall, when applying this threshold, we find that, on average, ${\sim}\npernightAlg{}$ of the objects submitted to the NEOCP per night \edtom{would require external follow-up}, but only around $\purityAlg{}\%$ (${\sim}\purityAlgRaw{}$) of those objects would be NEOs. Moreover, \edtom{by the end of the first year, $\neoLostAlg{}$ NEOs that were not prioritised for external follow-up due to this algorithm would remain undiscovered by LSST (i.e.\ there is no 15 night window with 3 nights of observations in the first year of observations)}. Though providing only a minimal improvement on the purity of the submitted objects, this method would reduce the follow-up need by a factor of two. In reality, we intend to submit all objects and their self-follow-up probabilities to the NEOCP, regardless of the value of the probability, such that the follow-up community can apply their own thresholds or sorting based on this additional parameter.

\section{Further mitigation strategies}\label{sec:further_strats}

The simulations discussed in Section~\ref{sec:current_impact} present a potentially bleak picture for follow-up, one in which less than 10\% of NEOCP submissions are due to true NEOs. Even after applying the algorithm we describe in Section~\ref{sec:mitigation}, the purity remains low despite improvements to the overall number of submissions. Given the limited telescope resources available, this is less than ideal. We therefore next look into a few simple prioritisation strategies that can increase the purity of the follow-up sample.

\subsection{Prioritize trailed sources}\label{sec:trailed_neos}

\begin{figure}[tb]
    \centering
    \includegraphics[width=\columnwidth]{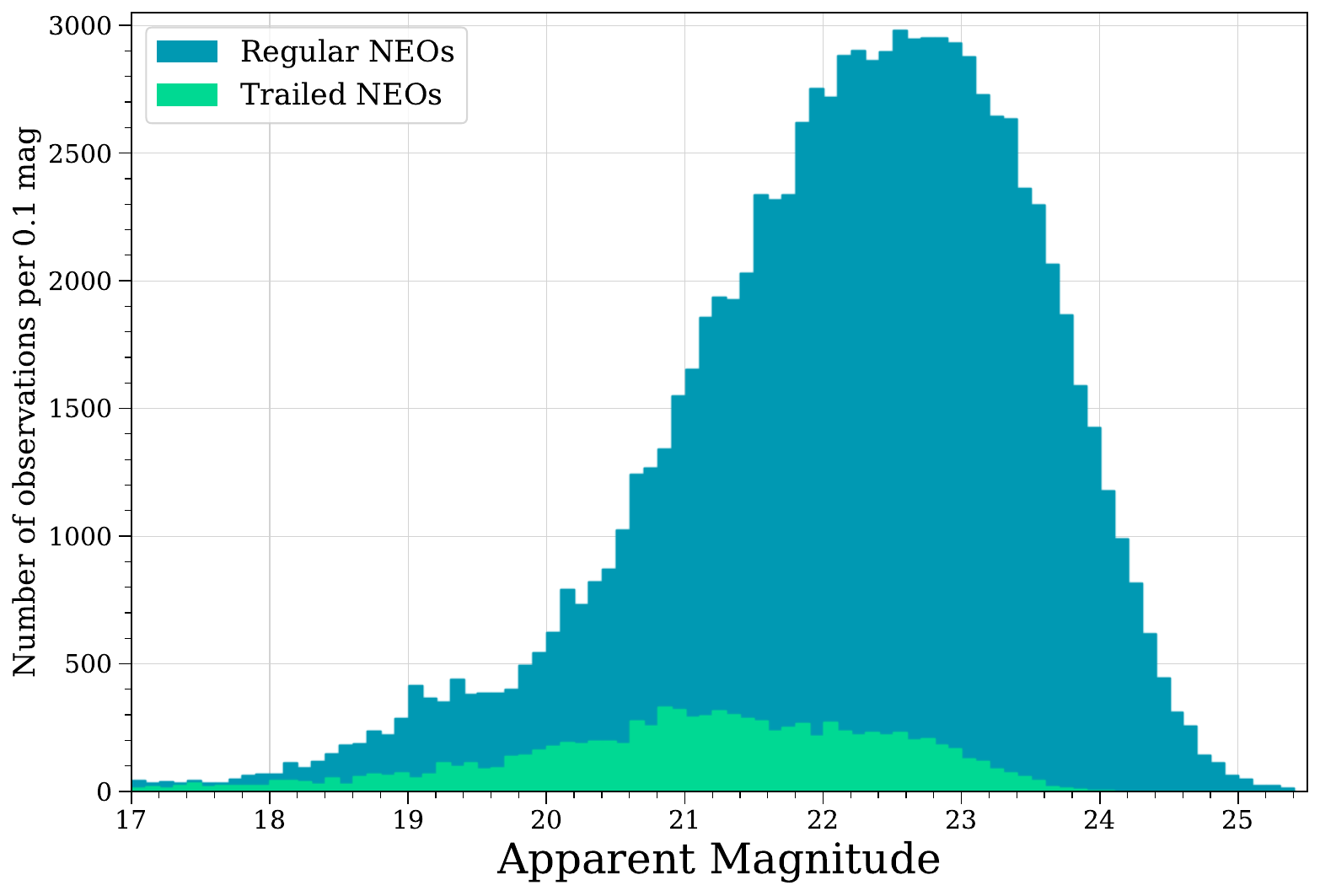}
    \caption{Apparent magnitude of LSST NEO observations in year 1, split up by whether they are trailed sources.}
    \label{fig:neo_magnitudes}
\end{figure}

A number of NEOs will be instantly recognisable as such due to their high angular velocities. An object moving faster than 1.5 degrees per day will leave trails of \edtom{at least 1.875} arcseconds in a single 30s LSST exposure, corresponding to just over 9 pixels (\edtom{1 LSSTCam pixel spans $0.2$ arcseconds on the sky}), which we adopt as our working definition of a trailed source. \edtom{Based on our simulations, LSST will, on average, observe 4 trailed NEOs in tracklets with at least 2 observations per night in the first year. No MBAs are trailed.}

Due to their ease of identification, trailed sources should be given priority in follow-up considerations. Their magnitude distribution is shown in Figure~\ref{fig:neo_magnitudes}. Approximately 72\% of trailed NEOs are brighter than $m_V < 22$; these may be especially good targets for the follow-up community to prioritise.


\subsection{Fainter sources are more likely to be NEOs}\label{sec:priority_mag}

Many objects observed by LSST will be too faint for external follow-up with presently available facilities. For example, the MPC observation archive\footnote{\url{https://minorplanetcenter.net/iau/ECS/MPCAT-OBS/MPCAT-OBS.html}} includes the magnitude of each observation at discovery. We cross-referenced this with the list of known NEOs and find that the 98th percentile in apparent V-band magnitude, $m$, is 21.73. For comparison, in Figure~\ref{fig:neo_magnitudes} we show the distribution of apparent V-band magnitudes of NEO observations in the first year of LSST. \edtom{We find that only around 40\% of non-trailed NEO observations in the first year of LSST are brighter than $m_V < 22$.}

\begin{figure}
    \centering
    \includegraphics[width=\columnwidth]{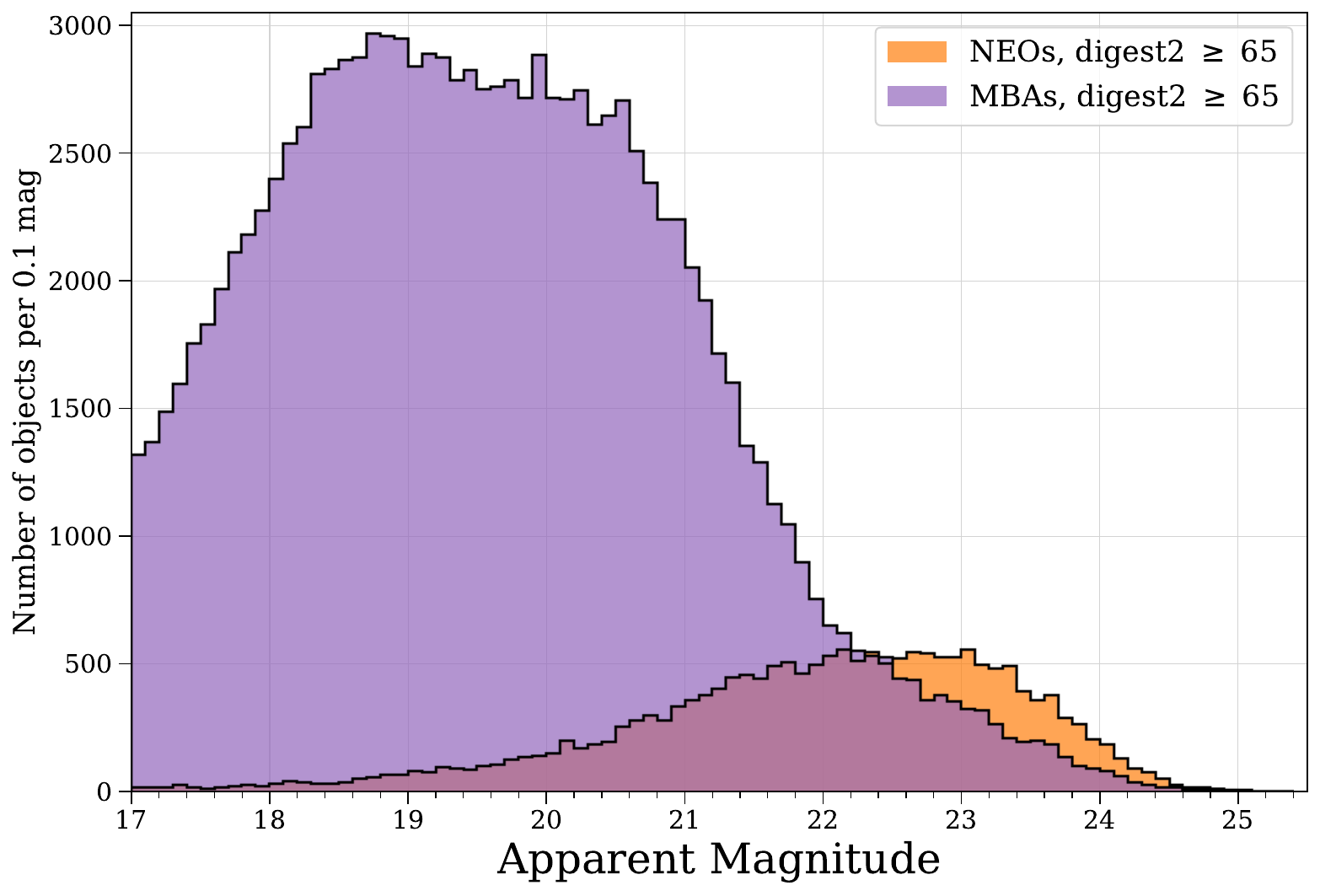}
    \caption{Apparent magnitude of NEOs and MBAs with a \dig{} score $\geq 65$ observed in the first year of LSST.}
    \label{fig:mag_after_digest2}
\end{figure}

However, if facilities have the capability to follow up fainter objects, this will improve their chances of finding NEOs. As shown in Figure~\ref{fig:mag_after_digest2}, we find that the likelihood of an object with a \dig{} score $\geq 65$ being an NEO increases significantly at fainter magnitudes. Indeed, we predict that 58\% of objects observed by LSST with $m_V > 22$ and a \dig{} score $\geq 65$ are NEOs.

The reason for the dearth of MBAs at fainter magnitudes is the \dig{} score cut. MBAs are assigned higher \dig{} scores if their orbits could be consistent with an NEO due to the degeneracy between distance and absolute magnitude (i.e.\ a source could be large and further away or smaller and close to Earth). For faint MBAs this is not possible because, for the already faint object to be closer to Earth, it would need to be incredibly small, which is not supported by the underlying population behind \dig{}. This leads to faint MBAs being removed by the \dig{} score cut.

Therefore, if facilities aim to maximise the chance of an object they follow-up being an NEO, they counterintuitively should prioritise \textit{fainter} objects\footnote{Note that this prioritises the \textit{number} of NEOs detections but would bias towards smaller NEOs, which would be less useful for missions focused on planetary defense.}. 

\subsection{Apply ecliptic latitude cuts}\label{sec:ecl_lat}

\begin{figure}[tb]
    \centering
    \includegraphics[width=\columnwidth]{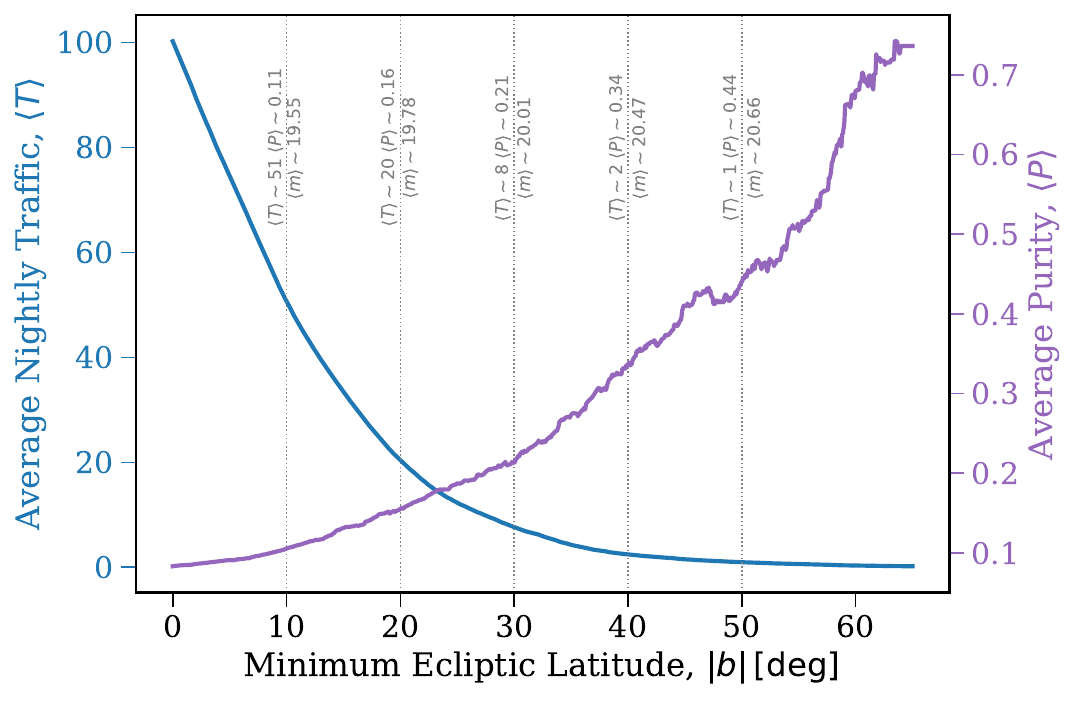}
    \caption{The blue curve shows the average nightly traffic of objects with \dig{}$\,\ge 65$ and apparent magnitude $< 22$ (left axis) and the purple curve shows the average purity (right axis) for a given latitude cut. Dotted grey lines at particular latitudes are annotated with y-axis values and average apparent magnitudes. \href{https://www.tomwagg.com/html/interact/neocp_ecliptic_latitude.html}{Interactive version}.}
    \label{fig:ecl_lat_cutoff}
\end{figure}

The final filter we examine in detail is applying an ecliptic latitude cut. Given that MBAs are constrained to lie within the ecliptic plane, ecliptic latitude, $b$, is a strong distinguishing factor between NEOs and MBAs. We explored the effect of applying this cut in Figure~\ref{fig:ecl_lat_cutoff}. We show, for a range of potential ecliptic latitude cuts, the resulting number of follow-up candidates (total number of NEOs and MBAs with \dig{} scores above 65) and average purity (fraction of those objects that are NEOs)\footnote{An interactive version of this plot is available online at \url{https://www.tomwagg.com/html/interact/neocp_ecliptic_latitude.html}}.

Inspection of this figure shows how adopting different $b$ thresholds is effective at increasing the purity of the resulting follow-up candidate list. For example, deciding to only observe objects at $b>30$ instantly raises the purity to $21\%$ though at the expense of completeness. Therefore, rather than applying cuts, the best strategy may be sort in descending order of ecliptic latitude when constructing follow-up lists. \edtom{We suggest that the MPC implement this feature on the NEOCP to allow users to easily prioritise in this manner}.


\subsection{Improve \dig{} algorithm}\label{sec:digest2_improve}
In Figure~\ref{fig:digest2_example} we highlighted that \dig{} struggles to distinguish between NEOs and MBAs when dealing with the large volumes of previously unknown MBAs that LSST will observe. For this reason, it is important to consider whether there are improvements that can be made to \dig{}. 

\paragraph{Population model} The population model used by \dig{} for orbit classification is S3M \citep{Keys+2019}. As noted in Section~\ref{sec:rescaling-s3m}, S3M does not accurately reflect our current knowledge of solar system objects. Therefore, updating the underlying population model, particularly after the first year of LSST observations, would help to improve \dig{}'s success in classifying NEOs.

\paragraph{Algorithm} Currently the \dig{} algorithm considers bins of perihelion, eccentricity, inclination and absolute magnitude to distinguish between different orbit classes. One could also consider other parameters for identifying NEOs, such as ecliptic latitude and the direction of motion on the sky. As noted in Section~\ref{sec:ecl_lat}, MBAs are constrained to reside within the asteroid belt and therefore close to the ecliptic. NEOs have no such restriction and hence observations with increased ecliptic latitudes are more likely to be NEOs. Moreover, since NEOs are much closer than MBAs, we would expect a much greater variation in direction of motion on the sky for NEOs. Specifically, it is unlikely for MBAs to be moving in a direction away from the ecliptic plane, whereas NEOs could be moving in almost any direction. In this way, the \dig{} algorithm could take into account further observational data, rather than classifying entirely based on the orbit, to more effectively classify potential NEOs.

\section{Summary \& Conclusions} \label{sec:conclusion}

We present new predictions for the NEOCP in the era of LSST. We performed mock LSST observations and used \dig{} to estimate the number of objects that LSST would send to the NEOCP using the current criteria. We created a new algorithm for predicting whether an object will be detected by LSST without external follow-up based on a single night of observations in order to reduce the load on the NEOCP. Our main conclusions can be summarised as:

\begin{enumerate}
    \item \textbf{Assuming no changes to listing criteria, LSST will significantly increase submissions to the NEOCP}\\Our simulations show that -- even though tracklets with $\ge 3$ observations occur only serendipitously within the baseline cadence -- LSST contributions will increase the nightly NEOCP submission rate by a factor of ${\sim}8$ over the first year to an average of \nightlyTrafficAllMags objects per night. Of these objects, ${\sim}\nightlyTrafficBase$ will be brighter than $m_V=22$. The rate will be the highest at the start of LSST and decline as more objects are identified. Due to the large number of tracklets due to an unrecognized MBA background in the first year of LSST, the fraction of submitted objects that are truly NEOs will be low, typically around ${\sim}\nightlyPurityBase$ (Section~\ref{sec:traffic_basic}).

    \item \textbf{Follow-up of objects that LSST itself will recover should be avoided}\\
    We construct an algorithm to identify submitted objects that are unlikely to require external follow-up. The application of this algorithm reduces the follow-up need by a factor of ${\sim}1.6$ (to ${\sim}\npernightAlg{}$ objects per night) on average.

    \item \textbf{Early NEO follow-up strategies should be highly selective}\\
    Selection on trailing, apparent magnitude, and additional scoring based on ecliptic latitude cuts can be used to prioritise the number of follow-up candidates. As the background of unknown MBAs diminishes with time, these constraints can be relaxed.
\end{enumerate}

\edtom{Finally, we note that the S3M simulations of \citet{Grav+2011} overestimate the number of solar system objects by about 25\% (which we've taken into account for our predictions; Section~\ref{sec:rescaling-s3m}).}

The purpose of this paper was to bring the impact of LSST on the NEOCP into focus and offer possible solutions. The increased number of candidates on NEOCP due to Rubin submission will make the (already arduous) job of prioritizing follow-up even more difficult. However, with the proposed mitigations and prioritization the number of candidates per night can be reduced. We therefore hope that, in anticipation of LSST's start in early 2025, the types of algorithmic mitigations we've developed here could be incorporated into tools such as NEOFixer \citep{NEOfixer}, the SNAPS broker \citep{SNAPS}, or other observation planning tools and aids. Also, the NEOCP inclusion criteria could be revisited and improved (e.g. inclusion thresholds, or stronger emphasis on sorting by \dig{} score), including improving the \dig{} score algorithm itself.

\clearpage


\begin{acknowledgements}
    \edtom{We are deeply grateful to the anonymous referee that provided detailed and helpful feedback for this work. In particular, we thank them for their comments that led to our realisation that S3M needed to be rescaled (Section~\ref{sec:rescaling-s3m}).} We thank Aren Heinze, Pedro Bernardenelli, Peter Vereš and Sarah Greenstreet for insightful discussions and feedback in many aspects of this project. TW thanks Andy Connolly and Stephen Portillo for their advice in assessing the quality of the hybrid catalogue, and Jessica Werk, Eric Agol, Željko Ivezić and Victoria Meadows for helpful feedback on an initial draft of this work.
    
    This material is based upon work supported in part by the National Science Foundation through Cooperative Agreement AST-1258333 and Cooperative Support Agreement AST-1202910 managed by the Association of Universities for Research in Astronomy (AURA), and the Department of Energy under Contract No. DE-AC02-76SF00515 with the SLAC National Accelerator Laboratory managed by Stanford University. Additional Rubin Observatory funding comes from private donations, grants to universities, and in-kind support from LSSTC Institutional Members.
    
    The authors further acknowledge the support from the University of Washington College of Arts and Sciences, Department of Astronomy, and the DiRAC Institute. The DiRAC Institute is supported through generous gifts from the Charles and Lisa Simonyi Fund for Arts and Sciences and the Washington Research Foundation. M. Juric wishes to acknowledge the support of the Washington Research Foundation Data Science Term Chair fund, and the University of Washington Provost’s Initiative in Data-Intensive Discovery.
    This work was facilitated through the use of advanced computational, storage, and networking infrastructure provided by the Hyak supercomputer system at the University of Washington.
\end{acknowledgements}

\software{\dig{} v0.19.2 \citep{Keys+2019}, \texttt{OpenOrb} \citep{Granvik+2009}, \texttt{difi} \citep{difi}, \texttt{THOR} \citep{Moeyens+2021}, \texttt{astropy} \citep{astropy:2013, astropy:2018, astropy:2022}, \texttt{matplotlib} \citep{Hunter:2007}, \texttt{numpy} \citep{numpy}, \texttt{pandas} \citep{mckinney-proc-scipy-2010, pandas_10045529}, \texttt{python} \citep{python}, \texttt{scipy} \citep{2020SciPy-NMeth, scipy_10909890}, \texttt{astroML} \citep{astroML, astroMLText}, \texttt{rubin\_sim} \citep{rubin_sim_10215451} and \texttt{scikit-learn} \citep{scikit-learn, sklearn_api, scikit-learn_10034229}.
This research has made use of NASA's Astrophysics Data System. Software citation information aggregated using \texttt{\href{https://www.tomwagg.com/software-citation-station/}{The Software Citation Station}} \citep{software-citation-station-paper, software-citation-station-zenodo}.}

\bibliographystyle{aasjournal}
\bibliography{paper, software}{}

\restartappendixnumbering

\allowdisplaybreaks
\appendix

\section{Historical NEOCP Submissions}\label{app:neocp}

The Near Earth Object Confirmation Page receives a varying number of submissions as a function of lunation and month and these totals have changed historically over the years. Historical archives of NEOCP submissions are maintained online by the Great Shefford Observatory \citep{neocp-archives}. We examine these archival records in order to assess how the submission rate varies. The collated machine-readable data and code for reproducing these plots is available on GitHub\footnote{\url{https://github.com/TomWagg/neocp-historical-trends}} and archived on Zenodo \citep{neocp-historical-code}.

\begin{figure}[htb]
    \centering
    \includegraphics[width=\columnwidth]{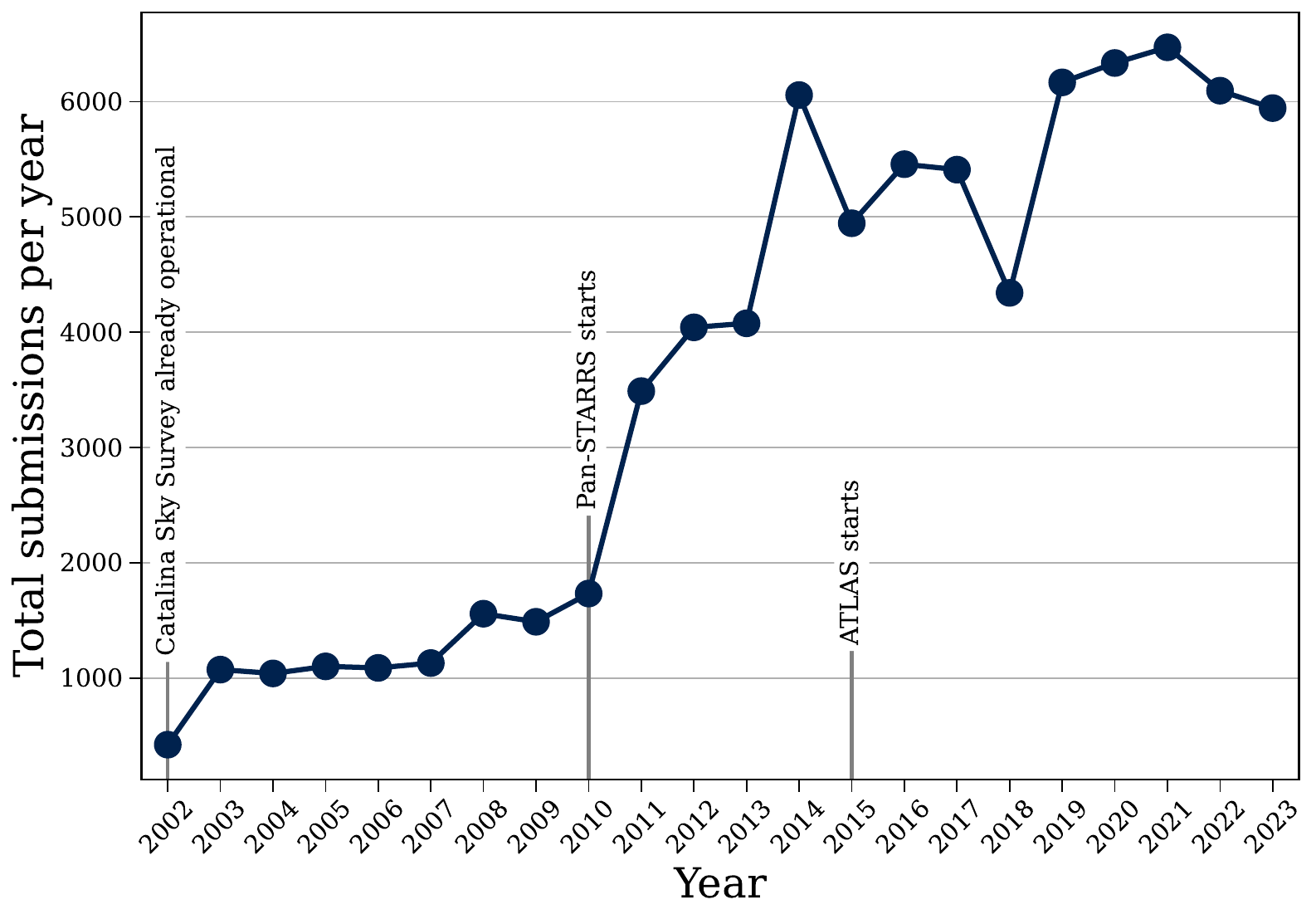}
    \caption{Total number of submissions to the NEOCP per year since 2002. Annotated lines show the start of operations of significant contributors to the NEOCP.}
    \label{fig:neocp-yearly}
\end{figure}

In Figure~\ref{fig:neocp-yearly} we show how the total number of submissions to the NEOCP has varied each year since 2002. The rate remained steady from 2002--2010, with primarily submissions from the Catalina Sky Survey. As Pan-STARRS1 begins operations in 2010 there is a sharp increase in the number of submissions which continues as Pan-STARRS2 and ATLAS start submitting observations. The rate over the past five years has remained relatively steady and thus we consider only these years when estimating the variations per month and lunation.

\begin{figure}
    \centering
    \includegraphics[width=\columnwidth]{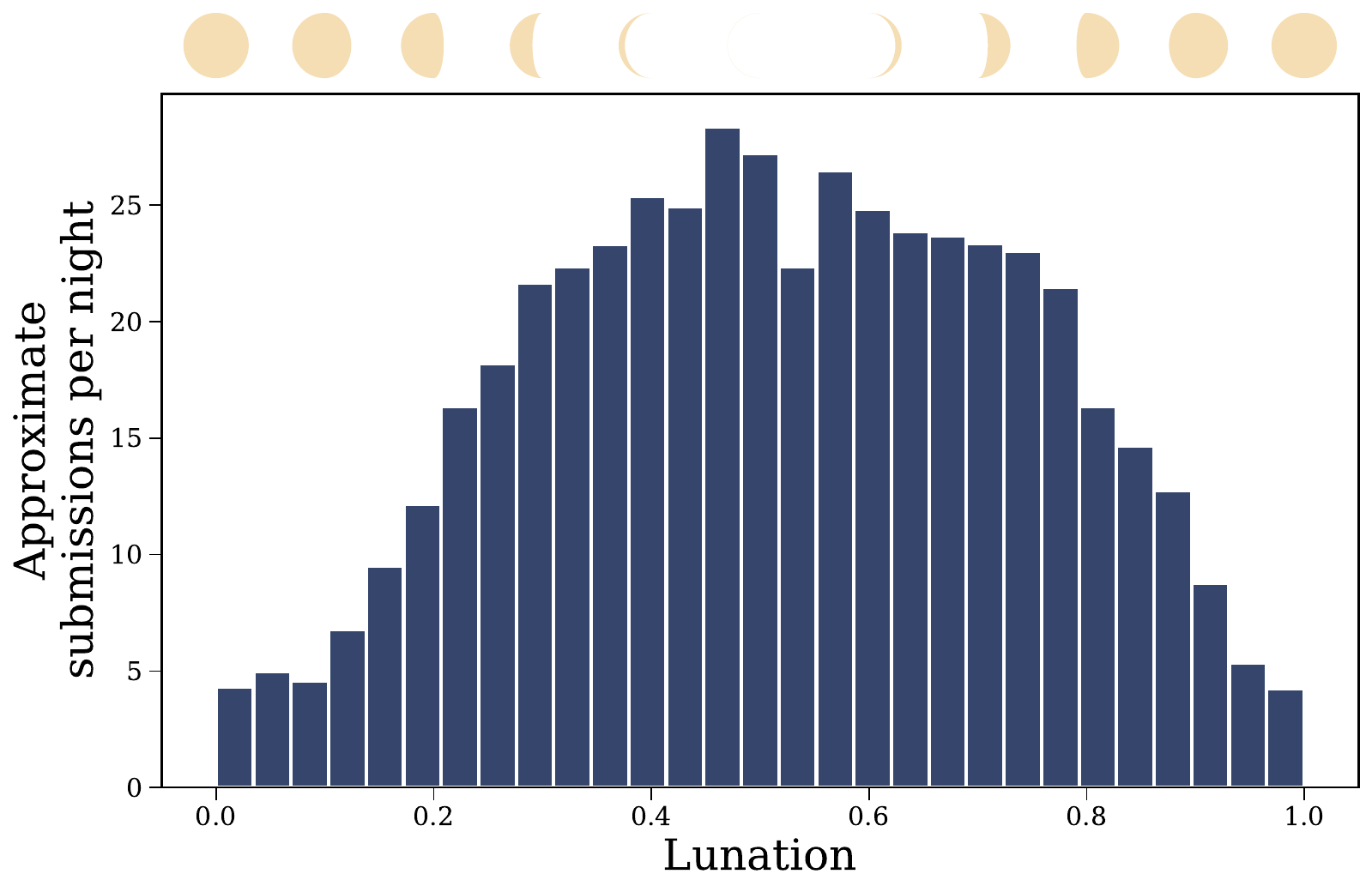}
    \caption{Nightly NEOCP submission rate as a function of lunation averaged over 2019--2023. Histogram shows 30 equally spaced bins over a full lunation, weighted such that each bin height corresponds approximately to an average nightly submission rate. Illustrations of the moon as a function of lunation are shown along the top axis.}
    \label{fig:neocp-lunation-overall}
\end{figure}

In order to calculate the rate as a function of lunation we use the \texttt{ephem} python package to convert each date from the NEOCP archive to a lunar phase\footnote{The dates in the archive are the MPEC dates and so there is a slight lag between this date and the submission date. However this lag is typically short and so this offers a good proxy for submission rate.}. We then calculate the number of submissions for each lunation over the past 5 years. In Figure~\ref{fig:neocp-lunation-overall} we show a histogram of these results, with 30 equally spaced bins over a full lunation, with each bin total dividing by the number of lunations in the past 5 years. This means that the height of each bin corresponds approximately to an average nightly submission rate. As one would expect, the number of submissions is at a minimum at each full moon, with around 5 submissions per night, and rises to a peak each new moon, with around 27 submissions per night.

\begin{figure}
    \centering
    \includegraphics[width=\columnwidth]{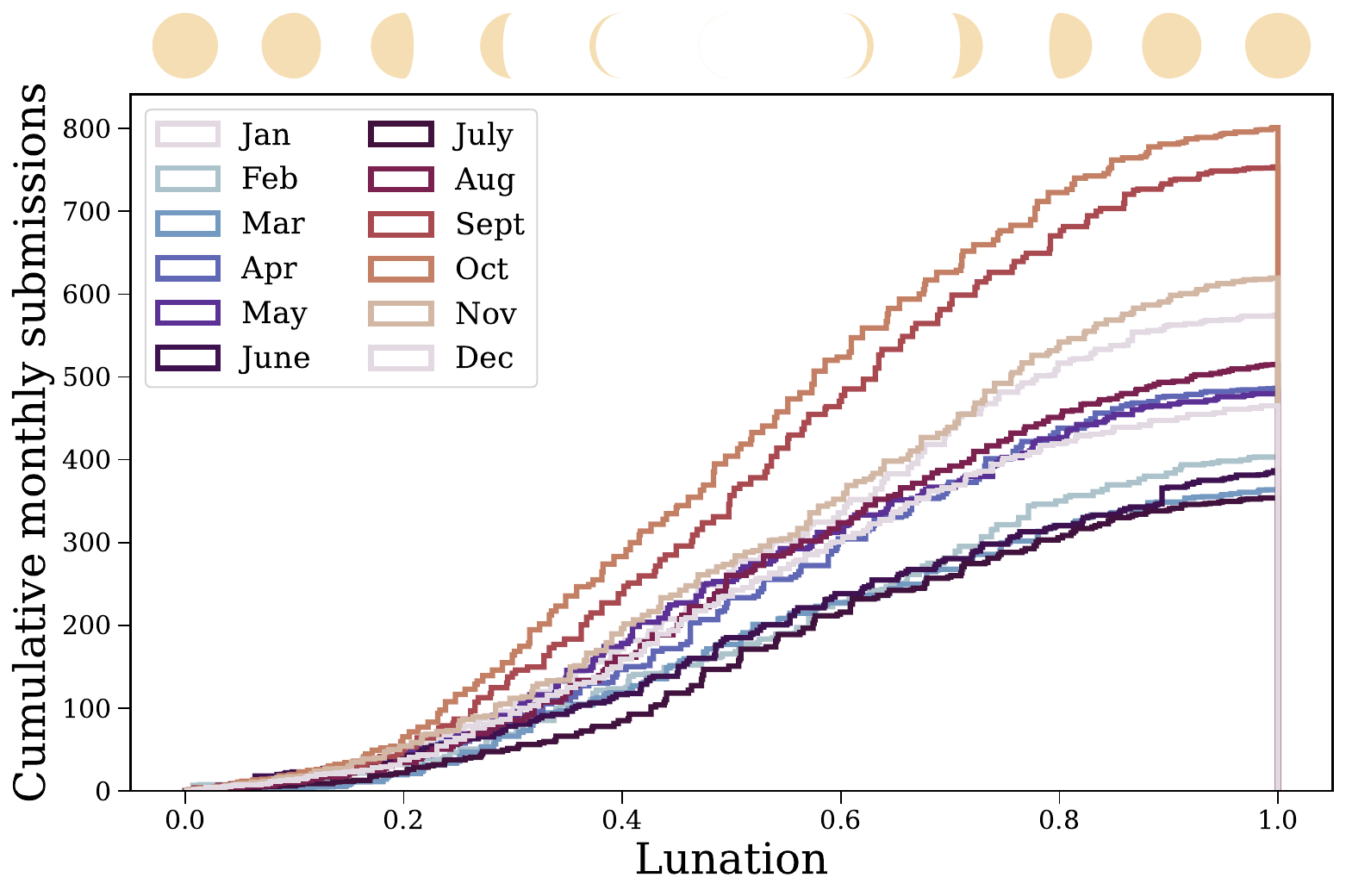}
    \caption{Cumulative montly submission rate to the NEOCP averaged over 2019--2023 as a function of lunation. Illustrations of the moon as a function of lunation are shown along the top axis.}
    \label{fig:neocp-lunation-monthly}
\end{figure}

In additions to variations by lunation, the rate also changes monthly due to seasonal variations. In Figure~\ref{fig:neocp-lunation-monthly} we show the cumulative number of monthly submissions as a function of lunation, averaged over 2019--2023. As in Figure~\ref{fig:neocp-lunation-overall}, the number of submissions close to a full moon is much lower than when there is a new moon. In addition in this figure one can note seasonal variations as the ecliptic plane moves through the sky (and due to changing weather patterns). For example, in March the total number of submissions only reaches around 320 on average, whereas in October the average total is instead around 800.

Overall we find that the number of submissions to the NEOCP during the era of LSST from other sources will be around 15 on average. This value will vary significantly over lunation and by month (and due to weather), such that on a given night there could be as many as 60 submissions or as few as none at all.

\section{Hybrid Catalogue Pipeline}\label{app:hybrid}
Many studies that make predictions for LSST use synthetic catalogue sof solar system objects that don't account for prior observations. In reality, we have already detected more than a million objects in the solar system and this number will continue to grow until LSST comes online. This means that current predictions of detection rates will be inflated, since a fraction of ``new'' detections may already be known. Therefore, for this paper we created ``hybrid'' catalogue that combines a synthetic catalogue with all known observations, whilst keeping the population distributions relatively unchanged.This

We created the hybrid catalogue to be dynamic, such that we can run a single pipeline to merge in an updated version of \mpco{} as more objects are discovered in the time until LSST comes online. All code to reproduce this hybrid catalogue is open-source and available on GitHub\footnote{\url{https://github.com/dirac-institute/hybrid_sso_catalogue}}.

\subsection{Data preprocessing}
For the synthetic catalogue of the solar system we use \sss{}, the Pan-STARRS Synthetic Solar System Model \citep[\sss{}][]{Grav+2011}. We merge this synthetic catalogue with the latest version of \mpco{} \citep{mpcorb}, a database of all currently known objects. We use \texttt{OpenOrb} \citep{Granvik+2009} to convert both catalogues to Cartesian coordinates and propagate all orbits until the same date.

\subsection{Merging algorithm}
The general idea for the merging algorithm is to inject each object from \mpco{} into \sss{}, replacing objects that are similar to those injected. An object's similarity is determined based on its position, $\va{x}$, velocity, $\va{v}$, and absolute magnitude (size), ${H}$.

\begin{figure}[tb]
    \centering
    \includegraphics[width=\columnwidth]{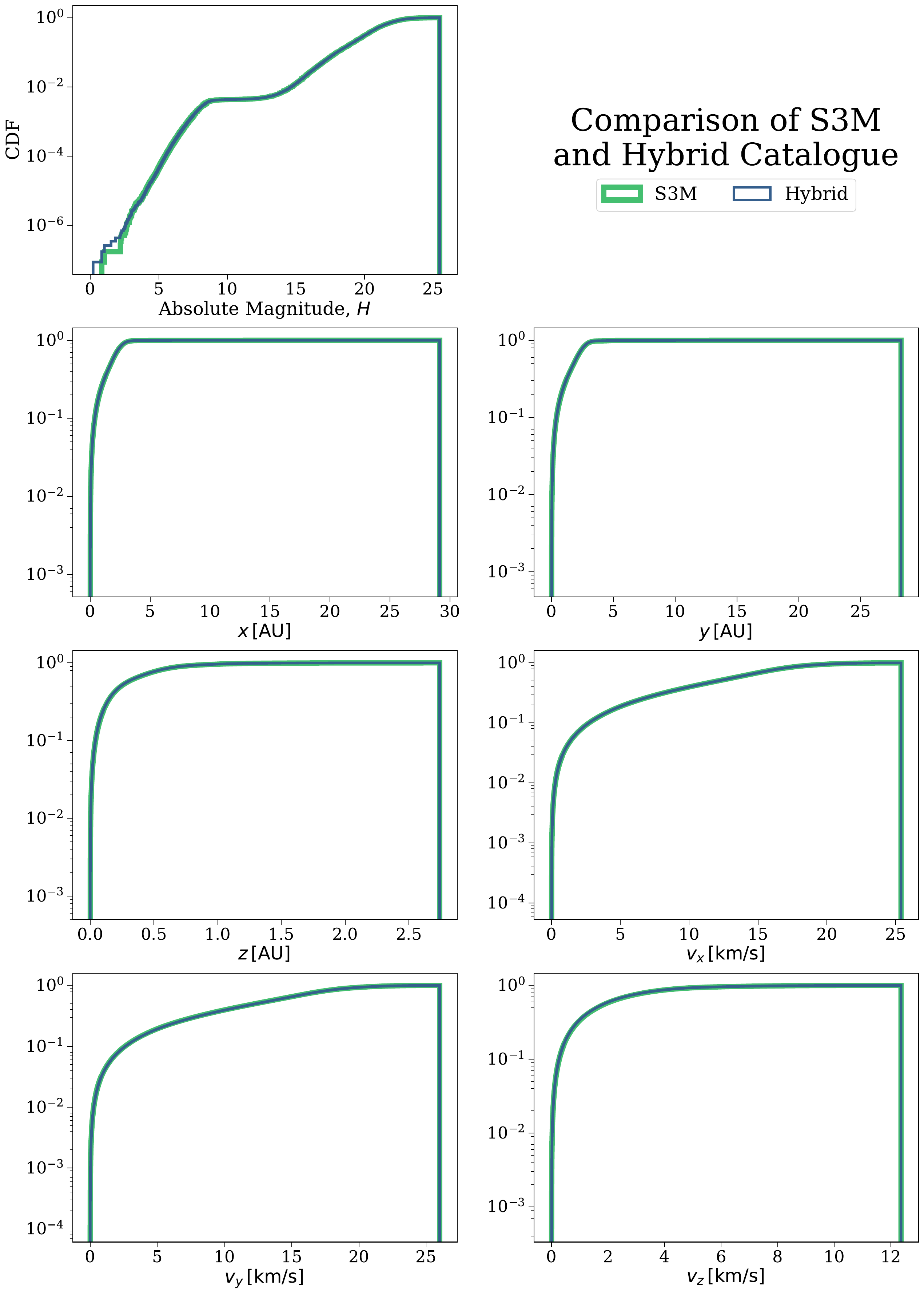}
    \caption{A comparison of the parameter distributions of \sss{} \citep{Grav+2011} and the hybrid catalogue we created.}
    \label{fig:hybrid_vs_s3m_dists}
\end{figure}

We split each catalogue into bins of absolute magnitude linearly spaced from $-2$ to $28$ and perform the merge algorithm on each bin separately. For each bin we build a K-D trees for both catalogues based on the positions ($x, y, z$) of objects. For every \mpco{} object we query the \sss{} tree for the nearest $100$ objects up to a maximum distance of $0.1 \unit{au}$, excluding any that have already been matched to a different real object. From these remaining nearest neighbours, we select the \sss{} object with the closest velocity as the matched object. If there were no remaining neighbours, either because no synthetic objects were nearby or because all nearby objects had already been matched, then we select a synthetic objects with the same absolute magnitude.

To complete the merging process, we compile the matched object IDs and delete them from \sss{}. We then add the entirety of \mpco{} to the remaining catalogue, resulting in a hybrid catalogue.

\subsection{Assessing quality of hybrid catalogue}\label{app:hybrid_quality}
It is essential that the underlying distributions of the hybrid catalogue do not differ significantly from \sss{} so that we still accurately reproduce the solar system. In Figure~\ref{fig:hybrid_vs_s3m_dists}, we show the distributions of the absolute magnitude and six orbital elements in both the hybrid catalogue and \sss{}. It is evident that the distributions are essentially identical, only showing slight differences for the brightest absolute magnitude objects (where S3M doesn't include simulated objects similar to those discovered in reality).

\begin{figure}[htb]
    \centering
    \includegraphics[width=\columnwidth]{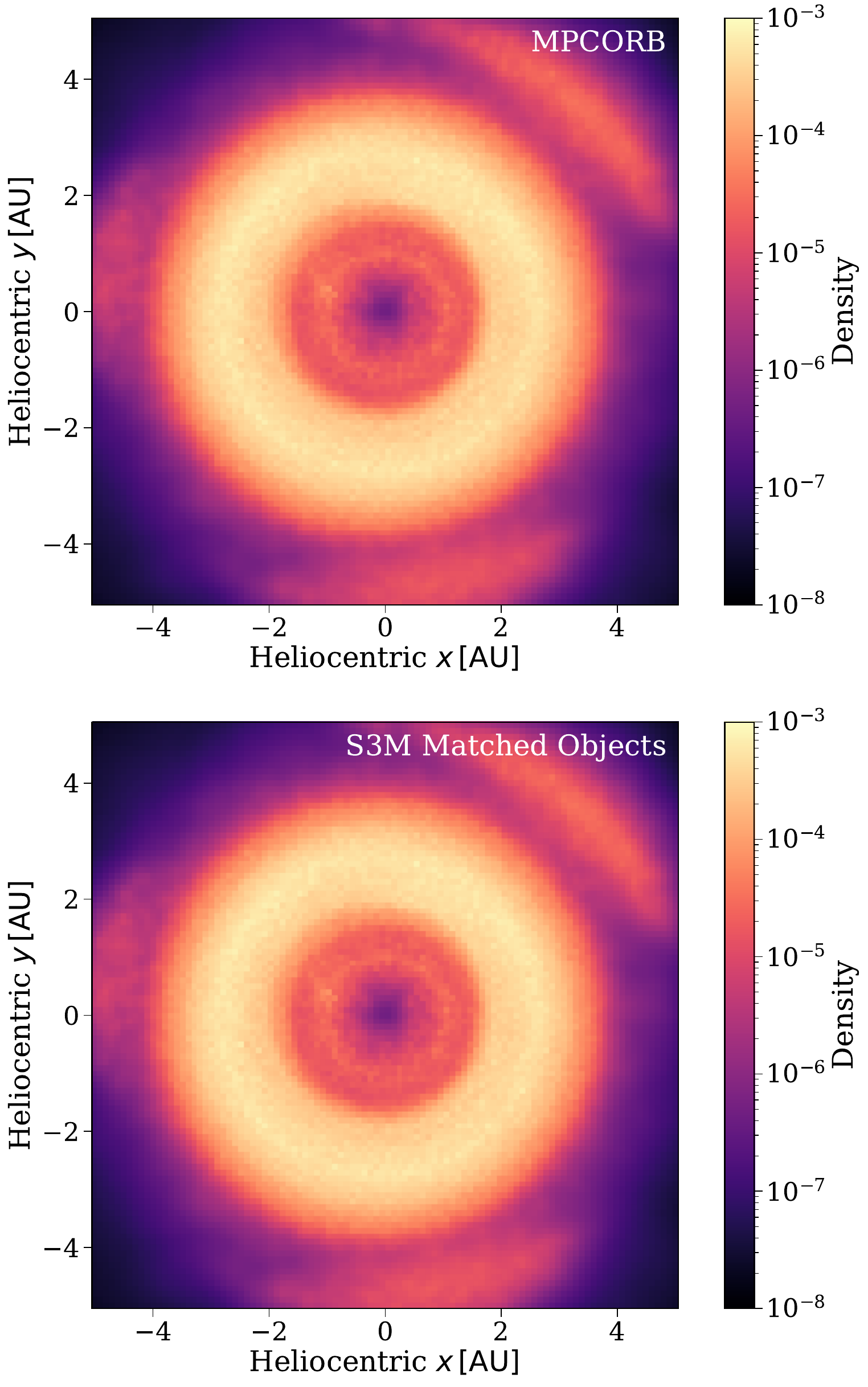}
    \includegraphics[width=\columnwidth]{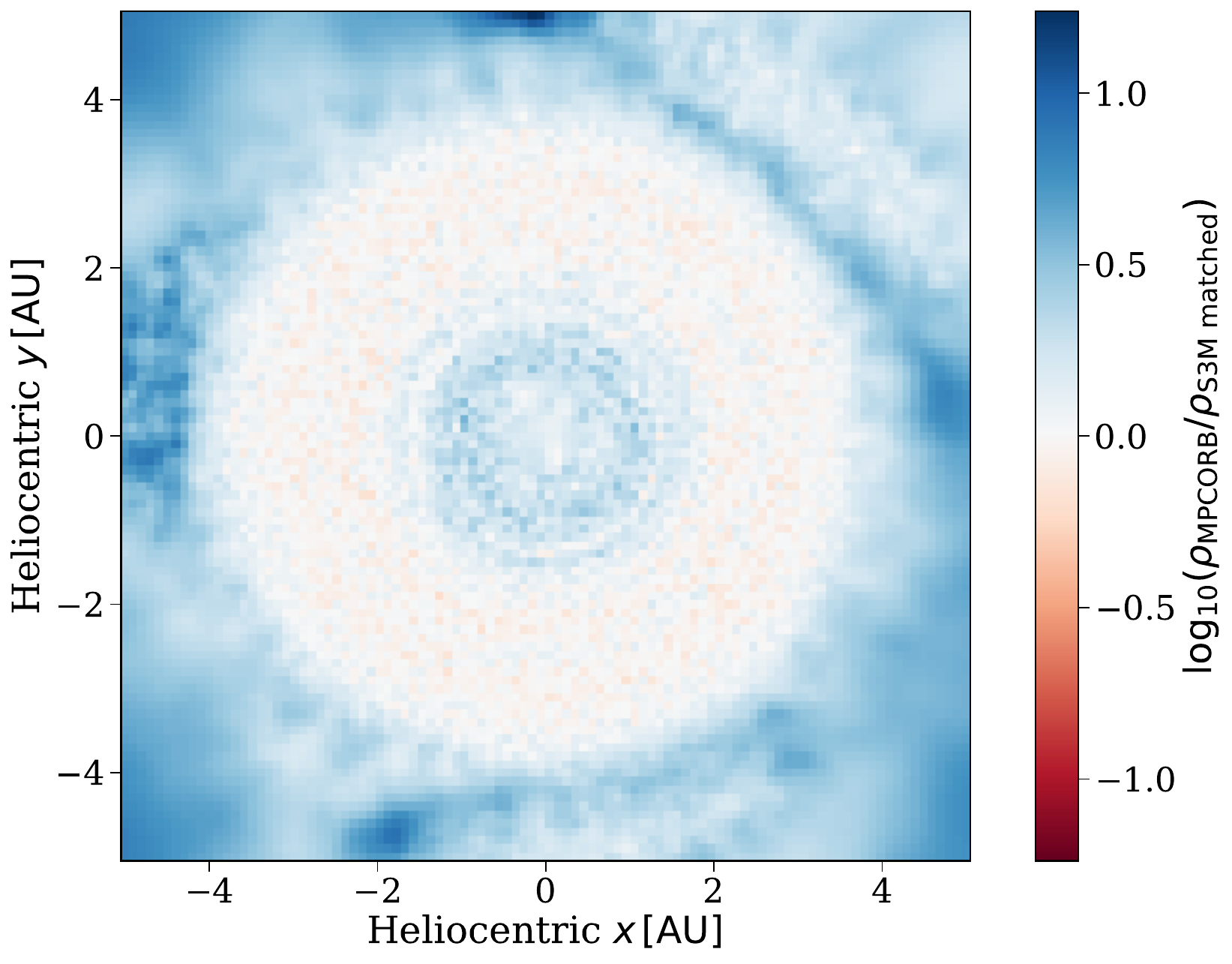}
    \caption{\textbf{Top two panels:} A comparison of the density of \mpco{} objects with those objects that were matched in \sss{} by our hybrid catalogue pipeline. \textbf{Bottom panel:} Residuals between the two plots above.}
    \label{fig:density_compare}
\end{figure}

As a further check, we compared \mpco{} to the objects that were removed from \sss{}, since these should have nearly identical distributions other than MPCORB objects that had no matches. In Figure~\ref{fig:density_compare}, we show a comparison of the densities for the heliocentric $x$ and $y$ and it is clear that these distributions are left unchanged in the hybrid catalogue.

\end{document}